\documentclass[preprint]{aastex62}
\usepackage{amsmath,amstext}

\newcommand{\Myr} {\mbox{$~\text{Myr}$}}
\newcommand{\AU} {\mbox{$~\text{AU}$}}
\newcommand{\pc} {\mbox{$~\text{pc}$}}

\newcommand{\Msun} {\mbox{$~M_{\odot}$}}

\usepackage{xcolor}

\graphicspath{{./}{figures/}}

\submitjournal{ApJ}

\shorttitle{Cradle(s) of the Sun}
\shortauthors{Pfalzner \& Vincke}

\begin{document}

\title{Cradle(s) of the Sun}


\author[0000-0002-5003-4714]{Susanne Pfalzner} 
\affiliation{J\"ulich Supercomputing Center, Forschungszentrum J\"ulich, 52428 J\"ulich, Germany}
\affiliation{Max-Planck-Institut f\"ur Radioastronomie, Auf dem H\"ugel 69, 53121 Bonn, Germany}

\email{s.pfalzner@fz-juelich.de}

\author[0000-0002-3796-2022]{Kirsten Vincke}
\affiliation{Max-Planck-Institut f\"ur Radioastronomie, Auf dem H\"ugel 69, 53121 Bonn, Germany}


\begin{abstract}
The Sun likely formed as part of a group of stars. A close stellar flyby by one of the solar siblings is probably responsible for the sharp outer edge in the solar system`s mass distribution. 
The frequency of such close flybys can be used to determine the likely type of
birth environment of the solar system. Young stellar groups develop very quickly, expanding
significantly within just a few Myr. Here we model this strong dynamical development of young stellar groups and determine the resulting close flyby history. We find that solar system equivalents are predominantly produced in areas with stellar densities in the range 
 5 $\times$ 10$^4$ pc$^{-3}<$ n$_{local} <$  2 $\times$ 10$^5$pc$^{-3}$.
Remarkably we find that only two very distinct types of stellar groups can be considered as serious contestants as the cradle of the Sun ---  
high-mass, extended associations ($M_c >$ 20 000 \Msun) and intermediate-mass mass,
compact clusters ($M_c <$ 3000 \Msun). Current day counterparts would be the association
NGC 2244 and the M44 cluster, respectively. In these two types of stellar groups, close flybys take place
at a sufficiently high rate, while not being too destructive either.  A final decision between these two remaining options will require incorporation of constraints from cosmo-chemical studies.
\end{abstract}


\section{Introduction} 
\label{sec:intro}

In recent years, first solar sibling candidates have been identified \citep{Ramirez:2014, Bobylev:2014,Liu:2016,Martinez:2016,Webb:2019}, which strengthens the earlier argument that the Sun was born as part of a group of stars \citep{Adams:2010}. There are several properties of the solar system that have been interpreted as indicators of the influence of this stellar group on the forming solar system: 
\begin{itemize}
\item First, the sharp outer edge of the solar system at about 30-35 au,
\item second, the orbits of trans-Neptunian objects like Sedna, 2012 VP$_{113}$ and 2015 TG387\footnote{This group of objects is in the following referred to as Sednoids.}, and 
\item third, the relatively high content of $^{26}$Al and $^{60}$Fe in chondritic meteorites.
\end{itemize}

Each of the solar system features has been used to constrain the properties of the solar system birth cluster \citep{Ida:2000, Kenyon:2004, Adams:2006, Spurzem:2009, Mitchell:2011, Li:2015, Jilkova:2015, Owen:2010, Dai:2018}.  In particular the presence of the short-lived radionuclides $^{60}$Fe and/or $^{26}$Al were used to estimate the number of stars/mass of the solar birth cluster \citep{Thrane:2006, Dauphas:2011, Adams:2014, Parker:2014, Lichtenberg:2016, Nicholson:2017}. These calculations assume that these short-lived radionuclides were incorporated after a supernova outburst in the solar birth cluster.  \cite{Adams:2010} deduced a lower limit of $N>$10$^3$ on the number of stars in the solar birth cluster, translating roughly to a cluster mass of $M_c> 500 \Msun$.  Recently \cite{Portegies:2019} puts a very specific constraint of $N$ = 2500 $\pm$ 300 stars and a radius of $r_{vir}$ = 0.75 $\pm$ 0.25 pc forward on a similar argument. In recent years the requirement of a supernova explosion in the close vicinity of the Sun has been questioned as alternatives for the origin of $^{60}$Fe and/or $^{26}$Al have been suggested \citep[for example][]{Wasserburg:2006,Gounelle:2015,Fujimoto:2018}. However, the debate is still ongoing \citep{Goodson:2016,Nicholson:2017,Boss:2017,Lugano:2018}, therefore, it is not clear in how far above limits hold. Interestingly, chemical considerations demanding the presence of strong far-ultraviolet radiation fields similarly require $ N >$ 4000 \citep{Lee:2008}. However, very massive clusters with $N > 10^5$ have been excluded as solar birth clusters, because the correspondingly large number of very high-mass stars would produce a too strong radiation field \citep{Hester:2004,Williams:2007}. 

Based on flyby simulations considering the eccentricity and inclination increase of the giant planets, \cite{Li:2015} constrained the upper value of the membership to $ N < $ 10 000 assuming that the stellar birth group was a long-lived cluster. Similarly \cite{Brasser:2006} used Sedna's orbit to provide a lower limits on the central cluster densities of $\rho_{central} > 10^3$\Msun\ pc$^{−3}$, whereas \cite{Schwamb:2010} put an upper limit at $10^5$\Msun\ pc$^{−3}$. 
They also assumed that the stellar birth group was a long-lived cluster. Similarly, the orbits of the extreme transneptunian orbits \citep{Kenyon:2004,Kobayashi:2005,Jilkova:2015,Pfalzner:2018a} have been discussed in the context of the solar birth cluster.

 
  These previous investigations mostly took the properties of specific young observed stellar groups, often the Orion nebular cluster (ONC), as guidance for their cluster model. It lies in the very nature of observations that they present only a snapshot in time. Therefore most previous simulations tried to reproduce the properties in a quasi-static manner. However, this treatment can only answer the question of whether a stellar group is currently producing solar systems analogues, but not what happened in the past or future. 
  
  Recently it has become increasingly clear that young stellar groups are highly dynamical and expand considerably in size during the first 5--10 Myr of their existence \citep{Pfalzner:2009,Kuhn:2019}.  Another finding, important in this context, is that young stellar groups come in two fairly distinct classes, often referred to as clusters and associations. These two groups show distinctively different expansion histories  during the first 10 years of their development \citep{Pfalzner:2009,Portegies:2010}. As we will see, this renders a limitation of the solar birth environment in terms of the number of stars alone less meaningful.
  
 The surrounding group of stars could have influenced the solar system either by gravitational interactions with or the radiation from the other stars. The radiation consists of large amounts of EUV and FUV radiation arising mainly from the massive stars \citep{Adams:2006,Winter:2018} along with X-rays that from more distributed sources within the cluster. Together they heat the disk which can eventually lead to its evaporation. For low-mass stars  ($\sim$ 0.5 \Msun) in the solar neighbourhood, external photo-evaporation potentially potentially destroys many of the discs surrounding them \citep{Winter:2020}. However, \citet{Adams:2004} and \citet{Winter:2020} also find that for solar-type stars ($M \sim 1\Msun\ )$ the situation is completely different, they are much less affected by external FUV photons. 
 Therefore, although these radiative effects might nevertheless be important in other environments, they are largely beyond the scope of this present work. Here we concentrate on the gravitational effects within the solar birth cluster, however, discuss this limitation in Section 4.2. 
 
The relative importance of the expansion history of clusters and associations in the context of the solar system birth environment was first considered in \citet{Pfalzner:2013}, which favoured massive associations as the solar birth environment. However, our previous study considered only very massive stellar groups ($>$ 10 000 \Msun), the  expansion phase of clusters was only treated by fitting the observed the cluster sizes as a function of cluster ages and all flybys were approximated as co-planar. Here we take advantage of the recent observational and numerical progress in the understanding of early cluster/association development to up-date our investigation of the solar birth environment. We thrive for a better representation of the real situation by including non-coplanar flybys, modelling the physical effects that lead to the temporal development of clusters/associations, performing the simulations for a larger range of cluster/association masses, and modelling the embedded, expansion and semi-equilibrium phase of the cluster development self-consistently.

In section 2 we describe the numerical method including the initial conditions, the dynamics for the different cluster/association models (section 2.1) and the determination of the effect of flybys (section 2.3). The results concerning the probabilities of the solar system forming in the various environments are presented in section 3 and 4. In section 5 these results are compared to previous work and implications for future research are outlined.  \\

\section{Method}
\label{sec:solar:method}
A two step approach is applied, where first the dynamics of young clusters and associations is modelled while recording the close fly-by history of each star.  In a second step, this history is used to determine the flybys leading to a solar system-like cut-off at 30-50 AU of the outer disc.

\subsection{Cluster simulations}
\label{sec:solar:method:cluster}

The results presented here are based on an extensive set of simulations of the dynamics of different young clusters and associations \citep{Vincke:2016,Pfalzner:2018_PR,Vincke:2018}. In these works we used the code Nbody6++GPU \citep{Aarseth:2003,Wang} to simulate the stellar dynamics of various types of clusters/associations. The model parameters (see Table 1) are chosen to match the observed cluster and association properties including the temporal development of both groups \citep{Pfalzner:2013b}. For each model observational equivalents are indicated: models A1-A3/A4 can be understood as representative of Mon R2, the ONC, and NGC 2244 or NGC 6611, respectively, whereas C1 and C2/C3 roughly correspond to younger counterparts of the clusters Praeceps and Westerlund 2 or NGC 3603.  The parameters of models C2 and C3, and A3 and A4 are identical apart from the duration of the embedded phase.
The parameters of the various stellar groups are not chosen to be perfect matches to these specific clusters, but we model specific characteristic stellar environments. To aid connecting those to real observed clusters we name stellar groups that can be seen as representatives for the specific group of similar environments. The idealization used for our model clusters are discussed in more detail in section 4.4.

\begin{table}[h!]
\caption{Set-up parameters of the modelled associations and clusters.}
\begin{tabular}[t]{l|rrrrrrrl}
\tablewidth{0pt}
\hline 
\hline
model     & $ N_{\text{stars}} $ & $N_{\text{sim}}$ & SFE   & $r_{\text{hm}}$  & $t_{\text{emb}}$ & $t_{\text{dyn}}$ & type& represents 
\\
\hline
&		&	                     &       & [pc]             & [Myr]            &       &  &  \\ \hline
A1  &  $1\,000$  & 308 & 0.3   & 1.3  & 2.0 & 0.67&A & MonR2  \\
A2  &  $4\,000$  & 94 & 0.3   & 1.3  & 2.0  & 0.33&A & ONC \\
A3  & $32\,000$  & 9  & 0.3   & 1.3  & 2.0  & 0.12&A & NGC 2244,\\
    &           &           &       &       &  &      &       &    NGC 6611 \\
A4  & $32\,000$  & 9  & 0.3   & 1.3  & 1.0  &0.12 &A & NGC 2244,\\ 
    &           &           &       &       & &      &       &    NGC 6611 \\
C1  & $ 4\,000 $ & 50 & 0.7   & 0.2  & 1.0  & 0.03&C     & M44    \\
C2  & $32\,000 $ & 10 & 0.7   & 0.2  & 1.0  & 0.01&C     & NGC 3603, \\
    &           &           &       &       & &       &       &    Westerlund 2 \\
C3  & $32\,000 $ & 10 & 0.7   & 0.2  & 0.0  & 0.01&C     & NGC 3603 \\
    &           &           &       &       & &       &       &    Westerlund 2 \\
\hline 
\hline 
\end{tabular}
	\vspace{0.5em}
	\tablecomments{Col.~1 gives the model name, $N_{\text{stars}}$ (Col.~2) is the number of stars in the model, $N_{\text{sim}}$ (Col.~3) the number of simulations in the simulation campaign, SFE (Col.~4) the star formation efficiency, $r_{\text{hm}}$ (Col.~5) the initial half-mass radius of the stellar and the gas component, $t_{\text{emb}}$ (Col.~6) the duration of the embedded phase, Col.~7 indicates the dynamical time scale, Col.~8 gives the type of stellar group, where A stands for association and C for cluster and Col.~9 a cluster that roughly serves as young equivalent.}
	\label{tab:set-up}
      \end{table}

Starting at the end of the star formation process\footnote{This means that the times given are not necessarily equivalent to the cluster age as the star formation phase is not covered.}, the stellar group is still embedded in the gas, which is modelled as a background potential.  The cluster half-mass radii of  $0.2\pc$ and  $1.3\pc$, respectively, correspond to those typically observed at the start of the expansion phase \citep{Pfalzner:2013b} for the two types of stellar groups. Equally the star formation efficiencies (SFEs) correspond to those typically observed in associations (0.3) \citep{Lada:2003} and clusters (0.7)
\citep{Rochau:2010,Henault:2012,Cottaar:2012,Krause:2016}. The stellar density distribution is modelled as a modified King profile for the stars and a corresponding Plummer profile for the gas which reflects the situation in observed clusters \citep{Espinoza:2009,Steinhausen:2014}. The two profiles have the same half-mass radius meaning that the Plummer core radius is given as $r_c = r_{\text{hm}}/1.305$ \citep{Plummer:1911}. The stellar masses were sampled from an initial mass function \citep{Kroupa:2002} with a lower limit of $0.08\Msun$ and an upper limit of $150\Msun$. We assume the clusters to be initially in virial equilibrium and chose the velocities following the corresponding Maxwellian distribution. 

The development of the gas component at the end of the star formation process is complex and would require detailed knowledge of the gas dispersal mechanism. Given the lack of this knowledge, we make the simplifying assumption that the gas distribution follows roughly the stellar distribution, but is somewhat flatter in the central area. The latter being motivated by the fact that, for example the ONC is nearly gas-free in its centre. A Plummer profile fulfils this requirement of being flatter in the centre than the King profile applied for the stars. Using a Plummer profile for the gas has the additional advantage that it has a fixed outer edge, whereas King profiles do not. 

The gas is expelled instantaneously at the times $t_{\text{emb}}$ specified in table 1. Assuming such instantaneous gas expulsion is justified as in all considered cases $t_{\text{gas}} < t_{\text{emb}} < 1\Myr$. As we will see, the effect of gas expulsion is less pronounced for clusters than for associations due to the difference in SFE. For associations, gas expulsion is the main reason for cluster expansion and up to 90\% of stars become unbound. Clusters also expand despite their high star formation efficiency, however, the underlying reason is the high number of stars ejected from the densest central cluster regions \citep{Pfalzner:2013b} rather than being the effect of gas expulsion.

Tracking the equivalent of approximately 500 000 stars is required to guarantees that the error in the average disc size is smaller than 3\%. In order to fulfill this requirement we perform a considerably larger number of runs, $N_{\text{sim}}$, for each cluster model than is usual. Only for the dense clusters C2 and C3 we perform fewer runs, because the number of interactions makes these simulations computationally expensive. The individual simulations are combined by determining the respective properties in the individual simulations and calculating their mean value.

Binaries were not included in the initial setup, also correctly tracked when formed by capturing events. This choice was motivated by the fact that parameter space for the flybys would increase dramatically for binaries. However, this point should be investigated in future studies in detail.

\subsection{Effect of a fly-by on the disc size}
\label{sec:modelling_the_effect_of_a_fly-by_on_the_disc_size}

In the post-processing step the recorded close flyby history is used (periastron distance, $r_{\text{peri}}$, and the mass ratio of encounter partners, $m_{12}=m_{2}/m_{1}$) to determine the effect on the discs. Here $m_{2}$ denotes the mass of the perturber star and  $m_{1}$ the mass of the central star. Each star is treated as initially being surrounded by a 200 AU-sized disc. Here, the actual initial size is of minor importance, as long as it is significantly larger than the 30-50 AU tested for. 
%
%
The disc size for every solar-type star after a flyby,	$R_{\text{d}}$, is determined based on the parameter studies by \cite{Breslau:2014} and \cite{Bhandare:2016}. The simulation results of these two studies can be approximated by simple functions of the form 
\begin{equation}
	R_{\text{d}} = 1.6 \cdot r_{\text{peri}}^{0.72} \cdot m_{21}^{-0.2}, 
	\label{eq:bhandare_disc_size}
      \end{equation}
where $r_{\text{disc}}$ and $r_{\text{peri}}$ are given in AU and $m_{21} = m_2/m_1$ is the ratio between the masses of the perturber star ($m_2$)to that of the host star ($m_1$). For the solar system $m_{1}$= 1\Msun\, therefore this simplifies to %
\begin{equation}
	R_{\text{d}} = 1.6 \cdot r_{\text{peri}}^{0.72} \cdot m_2^{-0.2}. 
	\label{eq:bhandare_disc_size}
      \end{equation}
The disc sizes given are those averaged over all possible inclination for such a flyby. Previous studies were often restricted to flybys between coplanar orbits \citep[for example,][]{Pfalzner:2013,Portegies:2015,Vincke:2015}. As such
the values given here are more realistic because they take the inclination between the stars into account. 

We define stars that have masses between $0.8-1.2\Msun$ as solar-type stars. These constitute roughly $5\%$ of the stellar population in the stellar groups modelled here. Among those stars we search for flybys that lead to post-flyby disc in the range 30 AU to 50 AU. In the following, these are referred to as solar system analogues (SSAs).  
Equation 2 shows that flybys have to be very close to lead to such disc sizes unless the perturbing star is of relatively high mass. For example, for an encounter with an equal mass star, $m_{2}$=1, the periastron distance has to be of the order of 50-80 au. By contrast, a 50\Msun\ star would have the same effect passing at a distance of 110-175 au. 

While using Eq. 2 now includes inclined flybys, it is still restricted to parabolic orbits. Including the eccentricity would mean extending the already extensive parameter space by at least an additional factor of ten. Instead, we record the eccentricity in the cluster simulations so that we can see in which cases this approximation is justified and where it fails.  We will discuss in section 4 in how far this approximation influences the obtained results. 

Besides, viscous spreading after flybys is neglected, meaning disc sizes can only become smaller by consecutive close flybys. If the disc still contains significant amounts of gas, viscous effects can lead to an increase in disc size over time \citep{Cuello:2019}. This process tends to smeared out the outer discs edge. However, in the case of the solar system we deal with a sharp outer edge. Thus that the flyby must have occurred either just shortly before gas dispersal from the disc or afterwards, or it had a very low viscosity. Thus a purely gravitational treatment seems to suffice here. 

There is a slight inconsistency in our treatment as we do not take into account the matter captured from the perturber star. However, the captured matter moves usually on extremely eccentric orbits with short periastron distances \citep{Umbreit:2005}, as such it does not significantly contribute to the disc size, also it might constitute an important part in the TNO population \citep{Pfalzner:2018a}.

\begin{figure}[t]
\includegraphics[width=0.9\textwidth]{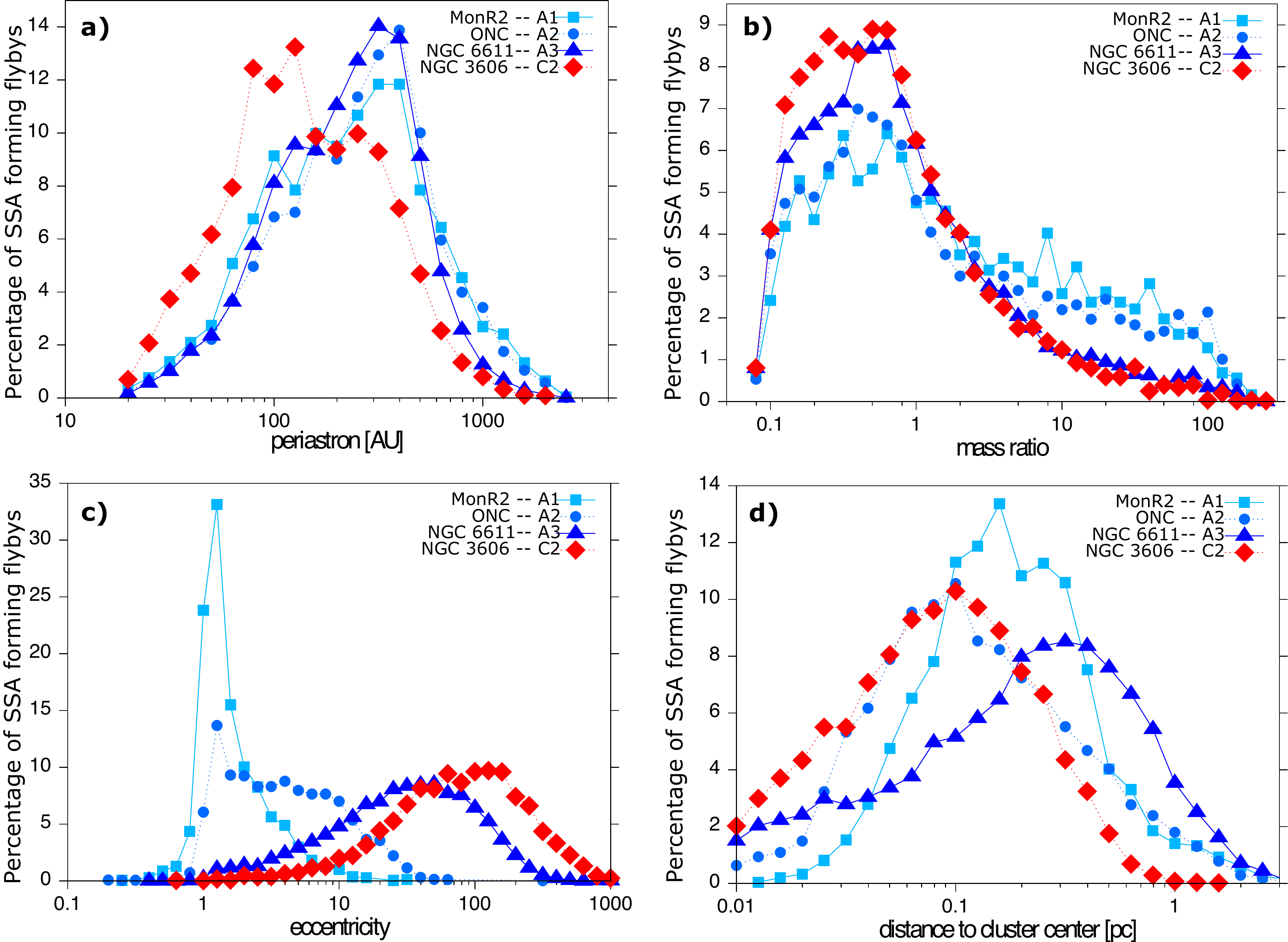}
\caption{Here only solar-type stars (0.8 \Msun $< m_1 <$ 1.2 \Msun) are considered that have a solar-system sized discs at the end of the simulation.  Shown are the percentage of stars where the solar system forming flyby had a specific property value. The parameters shown are a) the periastron distance, b) the mass ratio, c) the eccentricities and d) distance to the cluster centre. }
\label{fig:flyby_properties}
\end{figure}

 

\section{Results}
\label{sec:results}

First, we want to have a look at the properties of the flybys that lead to solar system analogues (SSA). Only solar system analogues (SSAs) are considered, meaning stars with masses between $0.8-1.2\Msun$ and post-flyby discs in the range 30 AU to 50 AU at the end of the simulation.
Figure~\ref{fig:flyby_properties} shows the percentage of stars that underwent a flyby with the respective properties: (a) periastron distances, (b) mass ratios, (c) eccentricities and (d) distances to the cluster centre of SSAs averaged over all simulations for models A1-A3 and C2, representative for clusters like Mon R2, ONC, NGC 6611 and NGC 3603, respectively. Models A4, C1 and C3 have been omitted, as they follow similar trends a A3 and C2. In a naive approach, one would expect that the distribution of periastron distances that lead to a 30-50 au-sized disc is independent of cluster environments. This is valid for the association models (A1-A3), however, in massive dense clusters (here illustrated for model C2) the SSA-forming flybys are on average closer with a maximum at 90 AU. 

Figure 1b gives some insight into the reason for this difference between clusters and associations. It shows the distribution of the masses of the perturbers that lead to SSAs.   While in all four shown cases the maximum of the distribution lies below 1, however, for model A1 and A2 (blue and turquoise line) a considerable proportion of perturbers have very high masses ($>$ 10 \Msun). As a result, on average,  the perturbers tend to have on average higher mass in the associations than in clusters (red line). The reason is that in associations the massive stars function as gravitational focuses for the low-mass stars in these comparably low-density environments and are largely responsible for the truncation of discs around low-mass stars. By contrast, in very compact clusters the stellar density is so high that the massive stars lose their dominant role as close flybys between low-mass stars become so common that gravitational focusing does not play such an important role anymore. The most massive stars are "shielded" by the high density of lower stars surrounding them. This means that here it is the equal mass-encounters that dominate the production of SSAs. Model A3 is basically at an in-between state because in the central areas the density is comparable to that in clusters but at the outskirts, it is more like lower mass associations. 


\begin{figure}[t]
\includegraphics[width=0.9\textwidth]{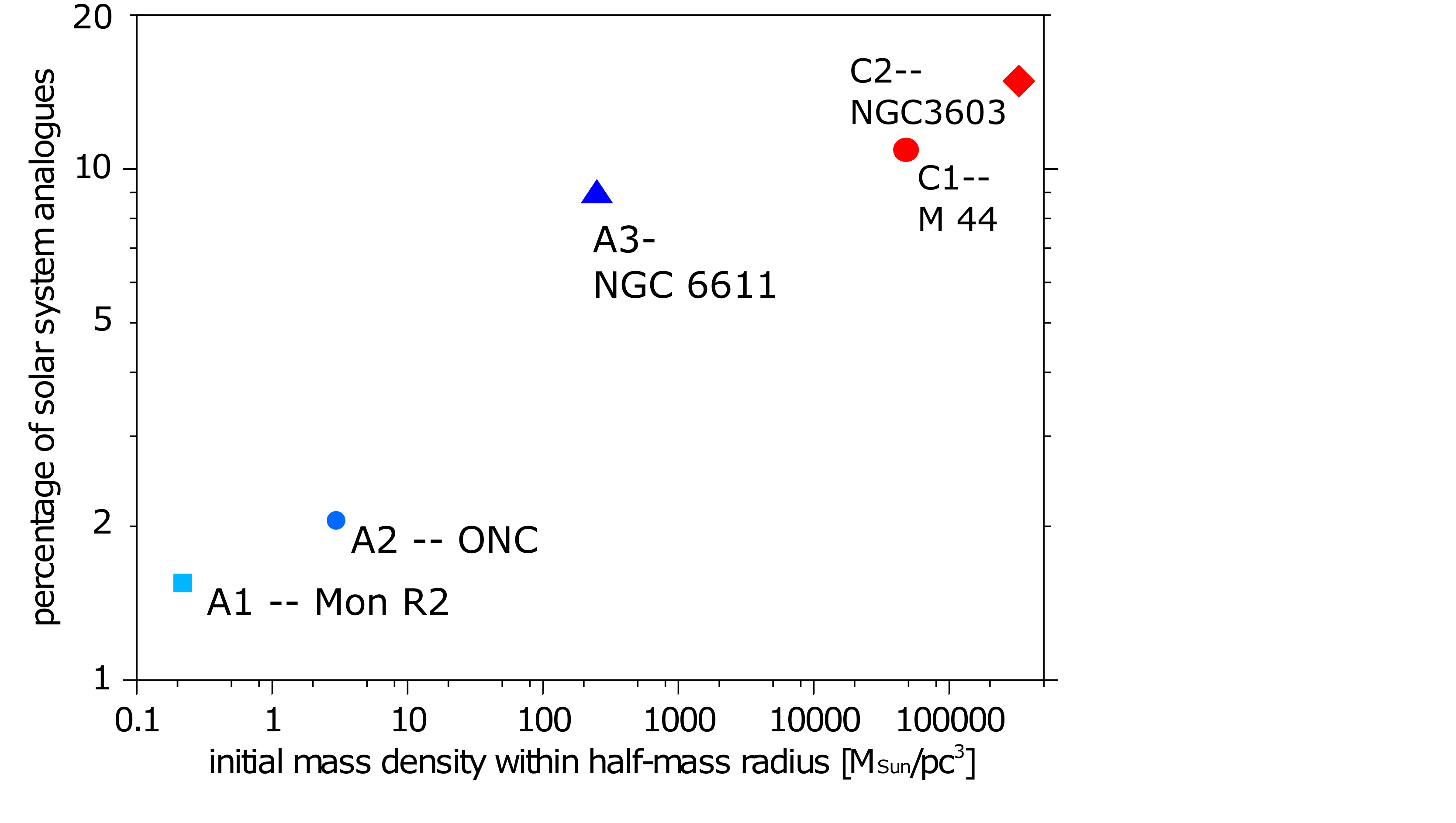}
\caption{Percentage of solar system analogues as a function of initial cluster density. Values are given for the time at the end of the simulation run.
}
\label{fig:density}
\end{figure}

This interpretation is strongly supported by Fig. 1c which shows the eccentricity distribution of SSA-forming flybys. For the lowest density associations, all flybys are on parabolic orbits ($e$=1), whereas for high-mass clusters the flybys show high eccentricities --- a similar behaviour was already noted in \citet{Olczak:2012}. However, hyperbolic flybys are expected to have a weaker influence on the disc \citep[see e.g.][]{Pfalzner:2004,Winter:2018}.  As we use the approximation of parabolic flybys in Eq. 2, we will discuss the consequences of this approximation on the results in section 4.1. 

Figure 1d shows the location of SSAs in terms of distance to the cluster centre at the end of the simulation. The position of SSAs varies strongly among the different association and cluster models and the explanation is somewhat more complex. For models A1-A3 the vast majority of SSAs are located within their half-mass radius of 1.3 pc. However, there are considerable differences in the location of the maxima in the distributions. Model A1 reacts slowest to the gas expulsion due to the low stellar density, therefore, most SSA forming flybys occur around the border to the cluster centre ($\approx 0.15\pc$). Model A3 is much dense, therefore expands quite violently after gas expulsion, while not dissolve completely. Therefore the centre is so dense that many close flybys lead to too small systems, therefore, the peak of the distribution is further out at approximately 0.4 pc. Model A2 is in-between in terms of density and dynamical reaction time. As a result,  most SSAs are very close to the association centre at about 0.1 pc or even less. It is pure coincidence that this matches more or less the peak in the distribution for model C1. For model C1 the situation is similar to A3, but more extreme in that most SSAs are located far from the actual centre at $\approx$ 0.3 pc, this means outside the initial half-mass radius of 0.2 pc. Outside of roughly $0.5\pc$, the number of SSA fly-bys becomes negligible. The reason is that the cluster expands comparatively little after gas expulsion because the SFE is so high.

%
%

\begin{figure}[t]
\includegraphics[width=0.9\textwidth]{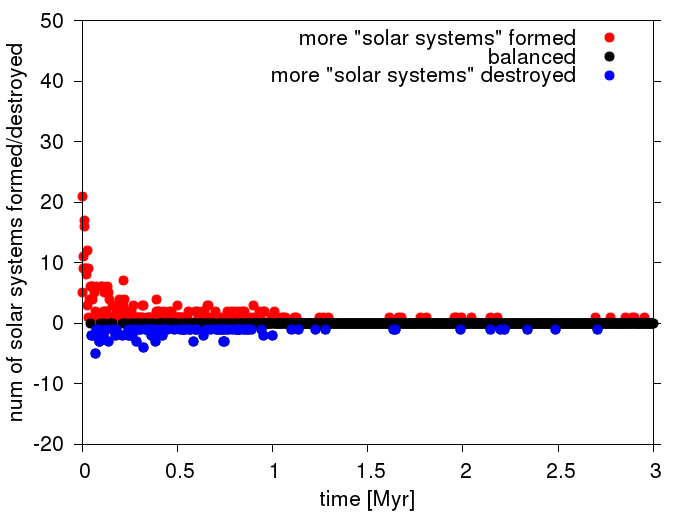}
\caption{Number of discs truncated within to SSAs (red) within a time interval of 0.01 Myr as a function of time and the number of SSAs truncated during that period to even smaller sizes, so that they are too small to form a SSA (blue). Here the case for model C2 is shown.}
\label{fig:formation}
\end{figure}

Figure \ref{fig:density} shows the percentage of SSA at the end of the respective simulation runs as a function of cluster density. Here density is defined as the initial mass-density (stars+gas) within the initial half-mass radius ($1.3\pc$ for associations, $0.2\pc$ for the cluster). In A1, but even the denser A2, the relative number of systems that are reduced to solar-system sizes is relatively low, only $1-2\%$ of the discs are solar system analogues. The latter value is in agreement with earlier studies in these environments \citep{Adams:2010}.

As to be expected the proportion of SSAs increases with density, and in much denser associations, like A3/A4, approximately 9-11\% of systems within the half-mass radius are SSAs. In compact clusters like C2, the portion increases even more to about 13 -17\% of stars, depending on model parameters.
However, it would be premature to conclude that clusters like C2 or even denser clusters are the most likely environment for the solar system to have formed in. Fig. \ref{fig:formation} shows the number of events per timestep of discs being reduced to SSA size (red circles) and those that are truncated below 30 au  (blue circles) as a function of time for the example of C2. with $R_d>$ 30 au, the latter systems are too small to form SSAs. In the initial stages ($<<$ 0.1 Myr) the number of SSA-forming events dominates over the destruction events, however, after just a short time the situation is reversed and there are much more SSA-destruction events than formation events.

\begin{figure}[t]
\includegraphics[width=1.0\textwidth]{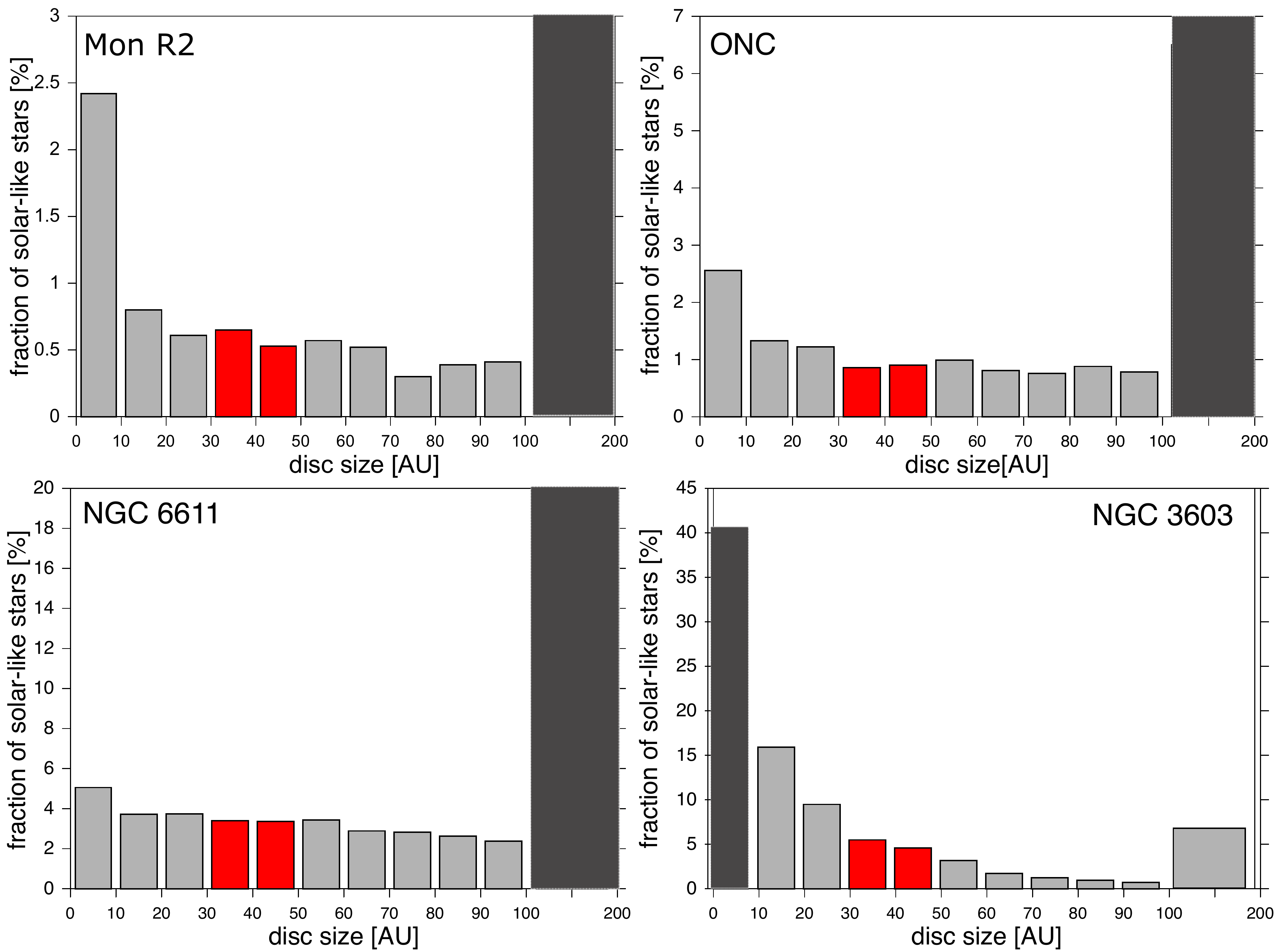}
\caption{System size distribution of simulated clusters A1, A2, A3 and C2, representative for Mon R2,  ONC, NGC 6611, and NGC 3603. The red bars highlight the system sizes similar to the Solar System, whereas dark gray indicates the most common disc size in the respective clusters.
Note that the last bin contains all discs having final sizes between $100\AU \leq r_{disc} < 200\AU$. The fractions shown are those for solar-type stars, meaning in the mass range 0.8 \Msun $< m_1 < $ 1.2\Msun.       
}
\label{fig:cluster_comparison}
\end{figure}

C2 is so dense that truncating to SSA-size is readily achieved, here the problem is that the systems do not become too small. This can be seen even better in the distribution of disc sizes given in Fig.\ref{fig:cluster_comparison}. Here the distribution is depicted at the end of the simulation runs for A1, A2, A3 and C2. It can be seen that for the associations, always the discs larger than the solar system dominant. 

In model A2, approximately 85\% of discs around solar-type stars are larger than $100\AU$, whereas roughly $5\%$ are smaller than 30 AU and only $\approx 1.8\%$ are solar-system-sized at an age of 10 Myr. In the less dense model A1, only $\approx 1.2\%$ of the systems resemble the solar system, while $\approx 95\%$ are larger than 100 au. The situation is different in the very massive association A3, where roughly $7-10\%$ of discs are SSAs and only $\approx$50\% larger than 100 au.

\begin{figure}[t]
\includegraphics[width=0.8\textwidth]{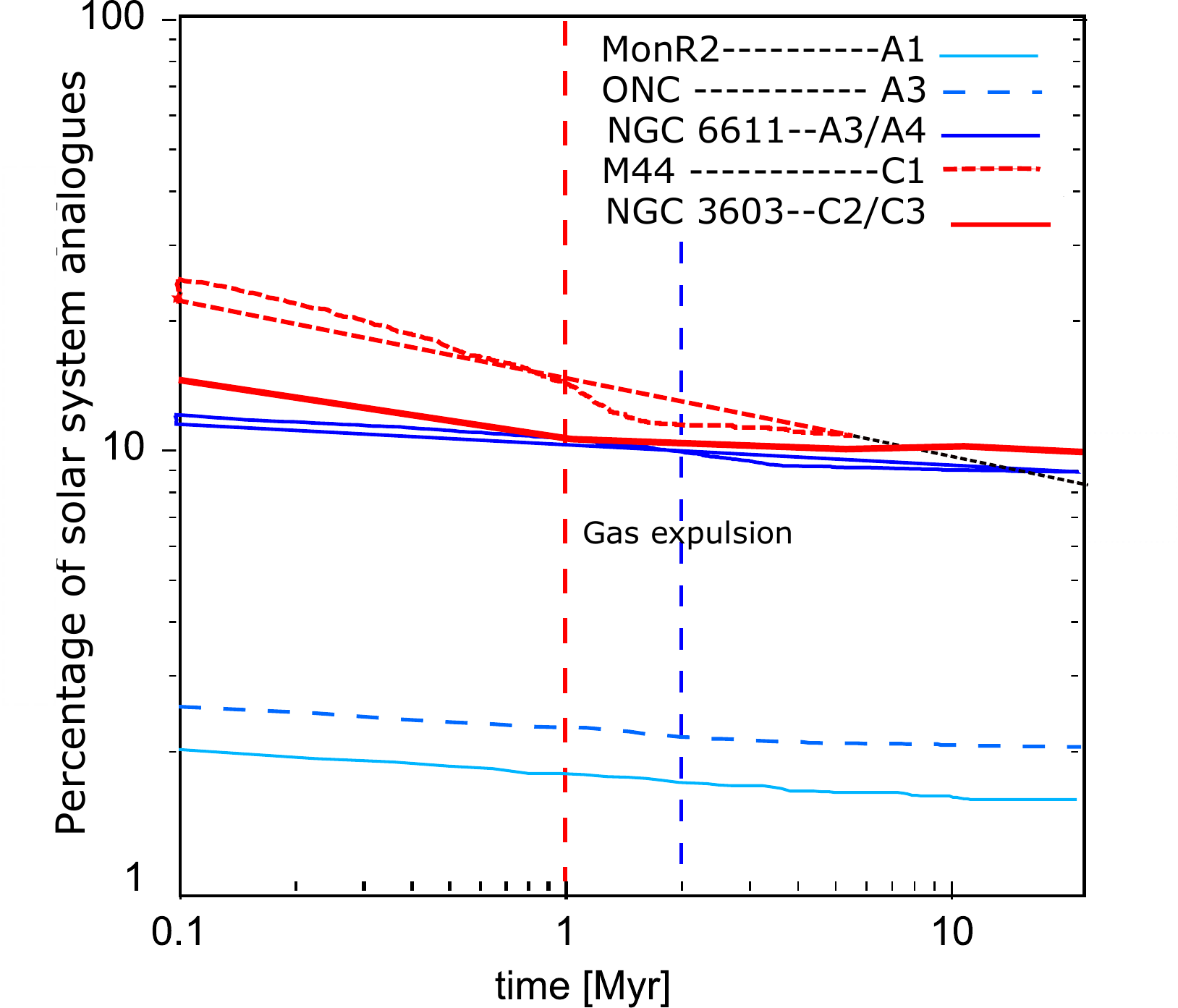}
\caption{Number of solar-system sized discs as a function of time for models A1-Taurus, A2-ONC, C1-M44, A3-NGC 6611 and C2-NGC 3603. Where there are two lines in the same line style, the models with a short and long embedded phase are both shown. The vertical lines indicate the time of gas expulsion as given in table 1.
}
\label{fig:time}
\end{figure}

The disc size distribution differs completely in dense clusters, where the average disc size is well-below 20 AU \citep{Vincke:2018}. Consequently, most discs ($\approx$ 66\%) are too small to form SSAs. Many of them ($40\%$) are $<10\AU$ implying that they contain so little mass that only very small planetary systems can form, if at all. Nevertheless, $ \sim 15\%$ of all solar-like stars are surrounded by SSAs. 

In absolute numbers, SSAs in "small" associations are rare. In A1 environments, on average less than one and, in A2-like associations, approximately four SSAs can be expected. The number of SSAs increases significantly in very massive associations,  with roughly 100-150 SSAs in A3 environments. For clusters like C2, it is even larger at about 150-250 SSAs.

However, these 15\% or 150-250 SSAs for C2-type clusters are representative only for $t$= 3 Myr. Figure \ref{fig:time} shows that the actual percentage of SSAs decreases on longer timescales. This effect is most pronounced in dense clusters. For model C3 already within the first 0.01 Myr, a large proportion of the discs are truncated to solar system sizes($\approx 50\%$). Afterwards, discs still experience SSA-forming flybys, however, an even larger number is "destroyed", i.e. truncated to $r_{disc} < 30\AU$. In summary, this leads to a constant decrease in SSAs. The situation is only slightly different when one considers a longer embedded phase (model C1). Here the different velocity distribution leads to initially slightly more SSAs as flybys that truncate the discs below 30 AU are somewhat rarer. However, this is out-balanced when gas expulsion happens, so that at 3 Myr the number of SSAs is nearly the same.

This decrease in the number of SSAs over time does also happen in systems A1-A3. However, the decline in the number of SSAs is always much slower than in clusters due to the lower density. It might be surprising that an overall decline does take place in these environments but the stars that encounter close flybys are usually located close to the cluster centre and often remain there so that the probability for a consecutive close flyby is higher for SSAs than for the population on the whole. 

This means that in dense clusters the problem is not that close flybys are rare but, in contrast, that they are so common. Discs are often truncated to smaller sizes than the solar system. As the cluster density changes very little after 3 Myr one can extrapolate this behaviour. At about 35 Myr the number of SSA in such dense clusters is lower in C2 than in A3, by then disc truncation to sizes $<$30 AU will have reduced the proportion of SSAs to less than 10\%. This destruction of SSA carries on until the clusters dissolve. As a consequence, it is more likely that the solar system formed in a cluster of smaller mass or a relatively high-mass association, like A3, than in a dense cluster like C2. \\ 

\section{Limitations} 
Two factors might influence our results, namely, the assumption of parabolic flybys and the restriction to gravitational interactions only. In the following we will discuss their possible effect.

\subsection{Parabolic flybys}

Above simulations assumed that all flybys were parabolic. As Fig. \ref{fig:flyby_properties}b illustrates, this is valid for Taurus-like stellar groups and to some degree also for ONC-like ones, so that our results should still hold here. However, the situation is different for NGC 6611, M44 and NGC 3603, where most SSA-forming flybys seem to happen on highly eccentric paths. The simplest solution seems to be to treat the flybys as hyperbolic. \cite{Olczak:2012} showed that hyperbolic flybys are usually less efficient in removing matter from the disc resulting in less mass loss and larger disc sizes. If this were the case, then for NGC 6611 and M44 assuming parabolic flybys probably overestimates the SSA-forming events to some degree, for NGC 3603 the situation is different because here the main hinderance for forming SSAs is that many discs are cut to smaller sizes. Therefore, here the number of SSA-forming destroying events might be overestimated. This could make particular a difference on the hundred to Gyr timescale.  

However, there is a problem in diagnosing the flyby paths in very dense systems. Implicitly it is assumed that, if no capturing process is taking place, the path is parabolic or hyperbolic and can be determined using three points of the orbit. This is valid if the problem can be approximated as a two-body problem and is a reasonably good approximation as long as the other stars of the cluster can be seen as perturbance to the two-body problem. However, in the very high-density regions, the actual orbit is neither parabolic nor hyperbolic, instead, the eccentricity of the orbit is constantly changing due to interactions with near neighbours. In this case, a three-point fit with the points being chosen before, at and after periastron passage, gives a fairly arbitrary value, as the property "eccentricity" becomes increasingly meaningless.  Therefore in these type of environments, a direct modelling of the effect of the discs would be the best approach. Unfortunately, these are the environments that are currently still not computationally accessible for such massive clusters.  


\subsection{External photo-evaporation} 
The proplyds in the centre of the ONC show that external photo-evaporation can reduce the sizes of discs, too \citep{ODell:1994,Ricci:2008}. In this study we did not include the effect of external photo-evaporation mainly because there is a large uncertainty on the mass loss rates and their temporal development and therefore on the timescales on which external photo-evaporation acts. Unfortunately, observation of the current mass loss rate in proplyds do not really resolve the issue. If these proplyds in the core of the ONC would have had the current mass-loss rate over the age of the ONC, \citep{Johnstone:1998,Henney:2002,ODell:2015} they should be destroyed by now, while in reality 80\% of stars are still surrounded by discs \citep{Winter:2019}. This is sometimes referred to as "proplyd lifetime problem".  Unlike for flybys,  the overall effectiveness of photo-evaporation seems to depend on parameters which so far are not well constrained. 

An important point in this context is that gravitational interactions and external photo-evaporation differ in the timing of their peak efficiency. Flybys are most efficient during cluster formation. By contrast, external photo-evaporation is least efficient during that period as the gas-dust density is so high that radiation is not very effective in penetrating. As gas-dust density decreases external photo-evaporation becomes increasingly important reaching its peak when gas expulsion sets in in the central regions of the stellar group. The ONC proplyds are an illustrative example of this phase. However, this phase where photo-evaporation is efficient last only for 1-2 Myr, afterwards the stellar density is too low due to the expansion as a reaction to gas expulsion. This sequence of events means that flybys set the scene in terms of disc sizes before photo-evaporation sets in.

The question is in how far do we expect external photo-evaporation to affect our results. 
Photo-evaporation is probably inefficient in Taurus-like environments as here both, the stellar density and the number of massive stars, is low. Therefore, the effect on the number of SSAs is probably negligible. In ONC-like stellar groups, photo-evaporation does truncate discs, but only in the clusters centre and only for a relatively short period. During the very early stages, the gas/dust density of the environment is so high that radiation is largely blocked and photo-evaporation is ineffective. At later stages and in particular, when the central area starts to becomes exposed during the gas-expulsion process, photo-evaporation can lead to the truncation of discs around stars that have massive stars in their neighbourhood. However, as soon as cluster expansion starts to set in at around 1 Myr, the association expands and external photo-ionisation becomes inefficient.

\begin{figure}[t]
\includegraphics[width=1.0\textwidth]{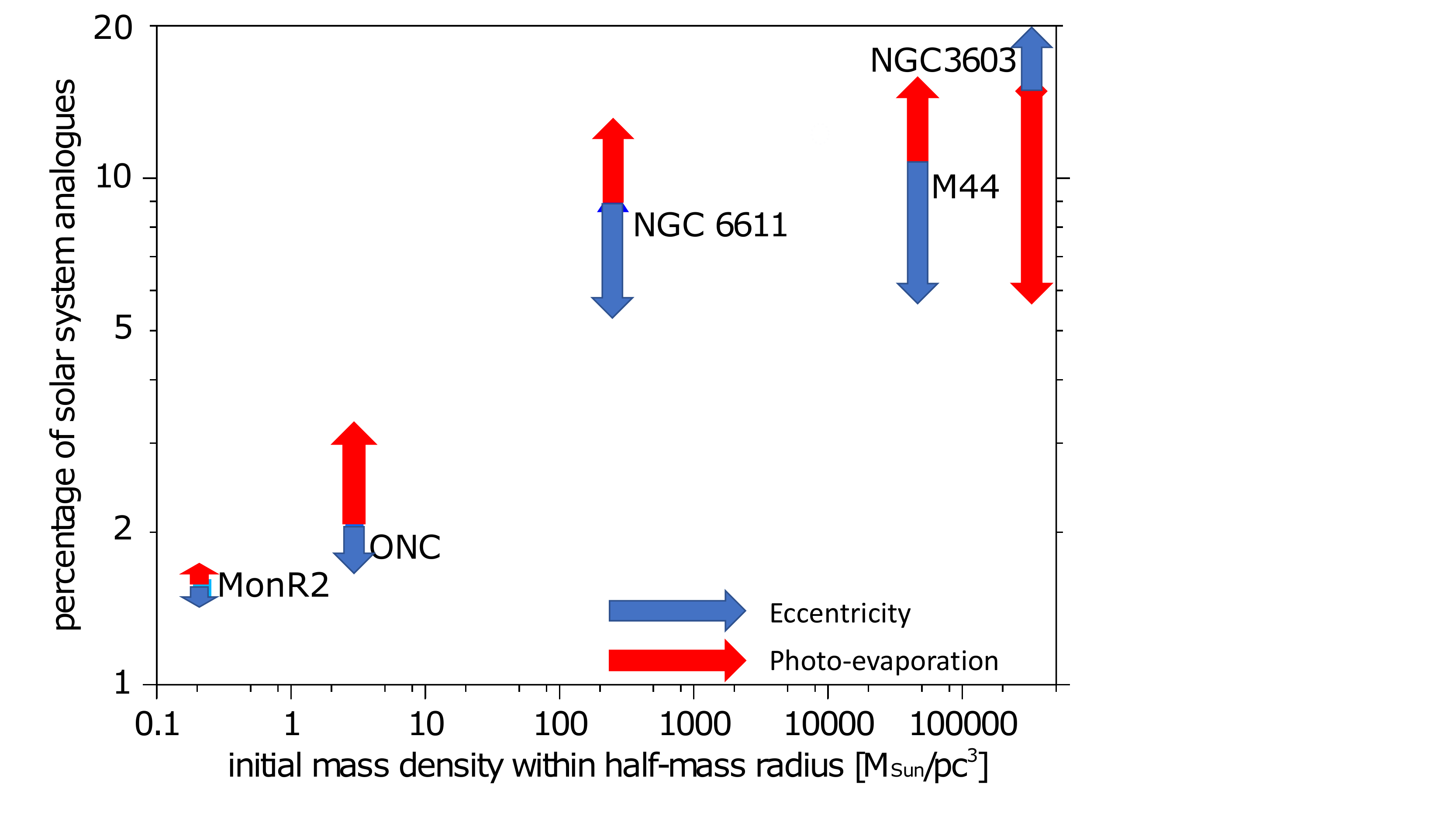}
\caption{Same as Fig. \ref{fig:density} but indicating the potential influence of external photo-evaporation and eccentrcity of the orbits on the percentage of SSAs for models A1, A2, A3, C2.}
\label{fig:uncertainty}
\end{figure}

The situation is quite different in clusters as they are much more compact and, although they expand over the first 5 Myr, most of their stars are still situated typically within just a few pc and most stars  remain bound. This means photo-evaporation is probably much longer at work in such an environment. For our densest cluster, NGC 3603, photo-evaporation is likely to decreases the number of SSAs as even more systems will be truncated to very small disc sizes. The situation is less obvious for M44, the lower density means that photo-evaporation could reduce large discs into SSAs, but also truncate SSAs to even making them too small.

Figure \ref{fig:uncertainty} tries to capture the situation including the above-discussed uncertainties. In summary, even considering the possibly effects of flybys and photo-evaporation, it is most likely that the solar system formed in relatively high mass association or a lower mass cluster. Only, if the effect hyperbolic encounters would dominate over that of photo-evaporation there would be a realistic chance that the solar system formed in a massive cluster like NGC 3603, Trumpler 14 or Westerlund 2. 

\subsection{Initial disc size}
We started out with the somewhat artificial assumption that all discs initially had a size of 200 AU. In reality, there will be a distribution of disc sizes. The actual distribution of disc sizes is not well constrained observationally, especially for the here relevant embedded phase.  However, it is probably realistic to expect that some stars are surrounded by a 30-50 au-sized discs at the time our simulation starts. 

One could argue that if the percentage of 30-50 au-sized discs would be sufficiently high, such disks would already be viable as solar-system analogues in terms of size and would change the percentage of successful systems. However,
for example Fig. \ref{fig:density} should not be interpreted merely as the percentage of stars that would be observed with a certain disc size. There is an important additional requirement to be a solar system analogue, namely, that the disc has a sharp outer edge like characteristic for the solar-system. This sharp outer edge is the main reason why it is assumed that a flyby is responsible for the size of the solar system.  Thus there might be 30-50 au-sized discs to start with in a real star cluster, however, those would not necessarily be a solar-system analog.

\subsection{Stellar groups}
The models of the various stellar groups are idealized in several ways, namely, as smooth centrally concentrated distribution with Maxwellian velocity distributions. Obviously none of the named equivalents fit these assumptions perfectly. Real stellar groups will deviate to different degrees form this idealization, by being  sub-structured and or elongated and their velocity distribution might deviate from the equilibrium implied by a Maxwellian distribution. For the compact and/or high-mass stellar groups (models A§/A4 and C1-C3) substructure is quickly erased and and a equilibrium situation established on a short time scale (see dynamical timescales in Table 1). This is confirmed by observations showing that the sub-structure leftover from the fractal structure of the parent molecular cloud vanishes typically within $\sim$ 1 Myr \citep{Jaehnig:2015}. For those environment above simulations should well represent the situation in real stellar groups. The situations is different for the lower mass associations, they might still show stronger signs from their formation process. However, neither elongation nor slightly lower velocities should lead to a significantly higher proportion of close flybys. By contrast,  sub-clustering might increase the number of close flybys. However, realistically it will not increase the number of close flybys in MonR2 or the ONC to such a degree to become comparable to those in NGC 6611, M44, or NGC 3603. In summary, these deviations of real clusters from our idealization should be inconsequential for our main result. 

\section{Discussion}

As mentioned in Section 1, there have been many previous constraints on the number of stars in the stellar birth environment based on different properties of the solar system. However, most did not distinguish between short- and long-lived clusters, respectively, clusters and associations. This limits the comparison that can be provided here. Despite, this difficulty, Taurus-like environments are generally regarded as very unlikely environments for the formation of the solar system, which we also confirm in our study.  Similarly, our results also agree with previous work that ONC-like stellar groups are only moderately likely solar system birth environments \citep{Lee:2008,Adams:2010}. 

On the high mass/number end, we agree with the works by \citet{Hester:2004} and \citet{Williams:2007} that very massive clusters like Westerlund 2 or Trumpler 14 are unlikely solar birth environments because here the density is so high and the radiation strong. However, an important point is that it is not primarily the high density and strong radiation that makes these environments too hostile for the solar system to have developed there, but it is mainly the fact that these high densities are maintained for at least 10s of Myr.

Although these findings agree with our results, we think that mass or number of stars alone are insufficient as a constraint.  One major reason is that there exist short- or long-lived clusters  (associations and clusters) so that one obtains different constraints on $N$ depending on which of the two different types of stellar groups one considers. But also distinguishing just between clusters and associations is not the best approach. In \cite{Pfalzner:2013} we considered both two types of stellar groups, even if it was with a number of approximations (see section 1). Our more detailed study confirms our earlier result that the close flyby frequency in NGC 6611-type clusters is higher than in the ONC.  Equally,  we come to the same conclusion that hardly any discs with solar system building potential would survive the initial highly interactive phase in very massive star clusters $M_c > 10^4 \Msun$. This makes high-mass dense clusters like NGC 3603 unlikely environments for SSAs. However, in \cite{Pfalzner:2013} we implicitly generalized this result that holds for high-mass dense clusters like NGC 3603 that to clusters in general.  Our new study shows that this is not valid. Intermediate-mass compact clusters similar to younger versions are actually a viable alternative to high-mass short-lived associations like NGC 6611 as solar birth environments.

As demonstrated neither cluster mass (number of stars) nor type of stellar group alone are good constraints for the solar birth environment. 
We will see that cluster density serves this purpose better but is also not ideal. Our simulations suggest that a cluster with an average peak cluster density of 100 pc$^{-3}$ to 10$^5$pc$^{-3}$ favour most is the formation of SSAs. The lower limit agrees with previous estimates by \cite{Brasser:2006}, however, our limit upper limit is higher than that of \citep{Schwamb:2010} which used the Sedna orbit as constraint.

Although more suitable than mass, the mean or central cluster density works only if one assumes that most SSA forming events happen in the cluster centre.  However, the frequency of SSA-forming events depends on the distribution of the stars. For example, stellar groups with Plummer profiles lead to lower close flyby rates than King profiles, equally, sub-clustering also alters the close flyby frequency.  We suggest using instead the proportion of stars that are located in the region where the stellar density favours SSA-forming events. 

Our simulations show that SSA forming events happen in 90\% of all cases in areas where the local stellar density exceeds $\rho_{local} > $ 5 $\times$ 10$^4$pc$^{-3}$. Similar SSA destroying events happen mostly where $\rho_{local}$ exceeds 2 $\times$ 10$^5$pc$^{-3}$. Therefore the proportion of stars that are contained in areas with 5 $\times$ 10$^4$pc$^{-3}<\rho_{local} <$  2 $\times$ 10$^5$pc$^{-3}$ gives a good indication  whether a certain environment is likely to produce many SSAs. Below this density SSA forming events are infrequent, whereas above SSA destruction becomes the norm. The advantage of using the local stellar density as a constraint is that these limits can be readily applied to any given distribution of stars. Therefore, it can also be applied to observed stellar groups to determine whether SSA forming events are likely to currently take place. 

Applying these constraints to our various models shows that the SSA forming flybys happen in completely different areas. Fig. 8 shows the proportion of stars that lie in this density band for the different stellar groups during the early phases of the evolution. To account for the different extent of the various stellar groups the location is shown relative to the initial half-mass radius of the various stellar groups and for comparison the also the initial mass density distribution in the cluster is shown. It can seen that for the ONC only a relatively narrow band close to the centre has the appropriate stellar density to produce SSAs. By contrast, the dominant SSA-producing region for NGC 6611-like systems and M44-like systems are not in the very centre but at a relatively broad region between 0.2 to 0.5 times the half-mass radius. In stark contrast, most SSAs in high-mass, NGC 3603-like clusters happen actually outside the half-mass radius. This means only if the solar system formed in the outskirts of such a cluster and became unbound within the first 5 Myr there would have been a chance to survive this harsh environment. 


\begin{figure}[t]
\includegraphics[width=0.6\textwidth]{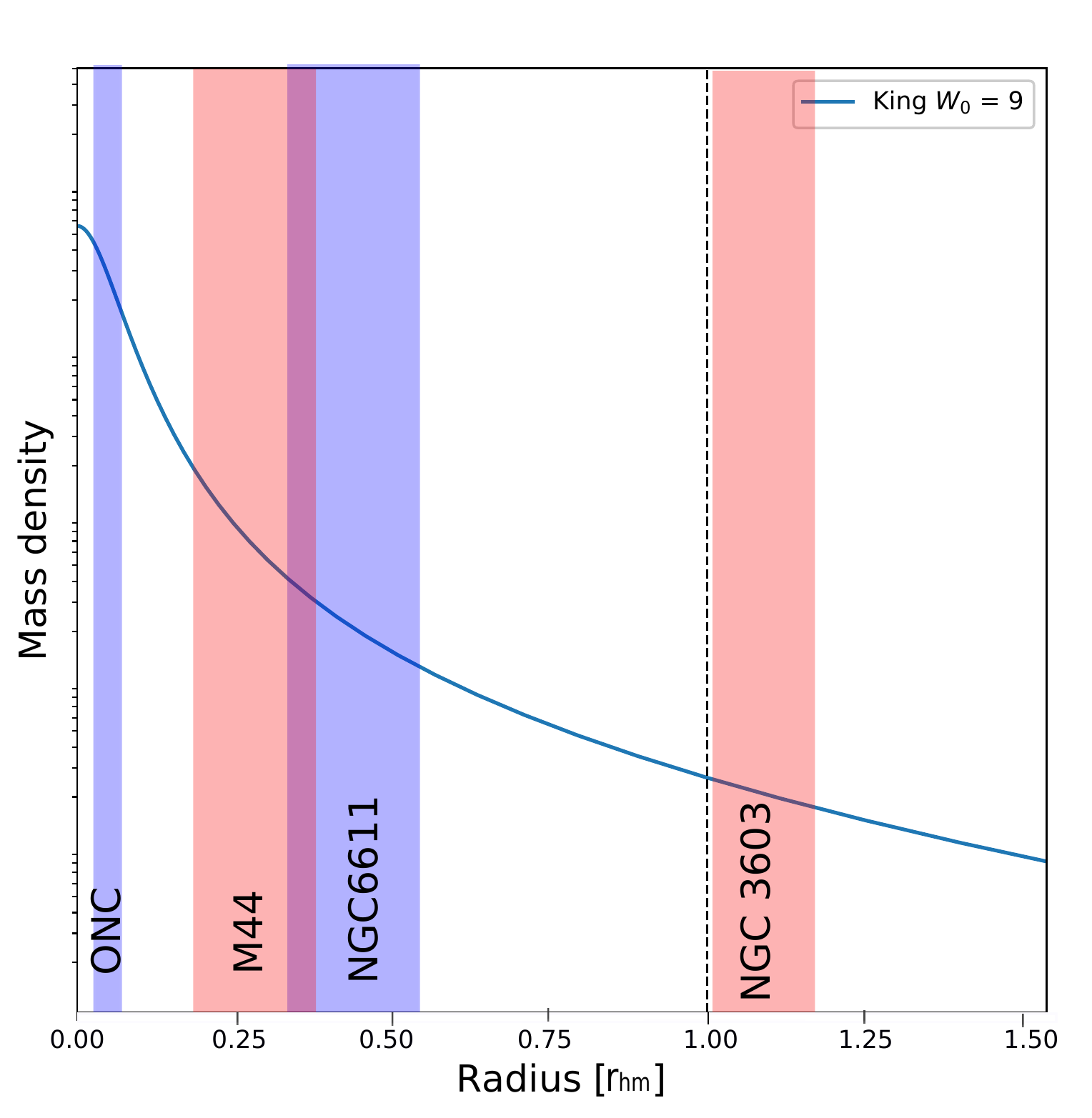}
\caption{The initial mass density distribution in the cluster models as a function of distance to the cluster centre normalized to the half-mass radius. The shaded areas indicate the regions most favourable for SSA production for the different cluster types.
}
\label{fig:active}
\end{figure}

\section{Conclusion}
\label{sec:conclusion}
Here we tried to constrain the type of stellar group that provided the birth environment of the solar system. 
Assuming the sharp outer edge of the solar system was caused by a close stellar flyby, we determined the frequency of solar system analogues being formed in different types of stellar groups. In contrast to most previous work, we explicitly take the temporal development of such stellar groups in account. Our simulations strongly suggest that the solar system was either born in a massive association ($M_c \approx$ 1-2 $\times 10^4$\Msun) or an intermediate-mass cluster ($M_c \approx$ 1-2 $\times 10^3$\Msun ), present-day equivalents being, for example, NGC~6611 and M44(Praeceps), respectively. In both environments approximately 10\% of solar-type stars experience such a solar system forming flyby and the formed system can be expected to remain intact over Gyr time scales afterwards. In future studies, above specifications of the two types of potential solar birth environments should be complemented by cosmo-chemical considerations. These could be used to come to a final decision about the solar birth environment.

Previously it has been investigated whether the Sun formed in a stellar group like the ONC. We find that the likelihood of solar system forming flybys is approximately a factor of 5 lower in ONC-like associations than those pinpointed above. The reason is that the peak density in the ONC is a factor 30-100 lower than in the above detailed environments. Similarly,  massive dense clusters ($M_c > $ 1 $\times 10^4$\Msun) seem also unsuitable environments for the formation of the Sun.  Although initially more solar system analogues are produced in such clusters, the systems formed are very susceptible to later destruction due to a high encounter frequency. The result is that most planetary systems in such environments can be expected to be much smaller than 30 AU. Only if the Sun formed well outside the half-mass radius  of such a cluster and became unbound fairly early on, it would have survived such harsh surroundings.

How can one determine whether a given stellar group is likely to form SSAs? We find that most SSA-forming flybys happen in the early phases at local stellar density in the range 5 $\times$ 10$^4$pc$^{-3} < n_{local} <$ 2 $\times$ 10$^5$pc$^{-3}$. In these environments the stellar density is sufficiently high to lead to a high number of close flybys, but also low enough to escape subsequent destruction.
Using the proportion of stars located at areas of such stellar densities provides a good constraint for determining the likelihood of forming solar system analogues.  is that these limits can be readily applied to any given distribution of stars. 
 
Above type of studies of the birth environment of the Sun depend heavily on understanding the dynamical development of young stellar groups. In recent years much progress has been made in the understanding of the general dynamics after cluster formation and the formation of low-mass clusters.   However,  the formation of clusters and associations with masses exceeding 1000 \Msun\ is less well understood. However, it is such massive stellar groups where the Sun most likely formed. Any observational progress in this field of the formation of high-mass stellar groups would help to determine the solar birth cluster even better.

Here only estimates of the long-term development of systems in the various environments were used, future simulations should cover the entire 4.5 Gyr of the lifetime of the solar system. This would provide better constraints on the number of stars being ejected and solar system analogues being destroyed.

This could mean that there is no real difference in the likelihood in this respect bewteen our two option, because more stars form in associations than in clusters, however, there are fewer high-mass than low-mass associations. Given the large uncertainties it could be equally likely that the Sun formed in any of the two remaining options.




\bibliographystyle{aasjournal}
\bibliography{references}


\end{document}